\documentclass[%
preprint,
showpacs,
 amsmath,amssymb,
 aps,
]{revtex4-1}
\usepackage{subfigure}
\usepackage{graphicx}
\usepackage{dcolumn}
\usepackage{bm}

\newcommand{\rmtwo}{I\hspace{-.1em}I}
\newcommand{\rmthr}{I\hspace{-.1em}I\hspace{-1.em}I}

\newcommand{\ti}{\tilde}

\begin{document}


\title{Entropy production of a steady-growth cell with catalytic reactions}

\author{Yusuke Himeoka and Kunihiko Kaneko}
\affiliation{Department of Basic Science, University of Tokyo, Komaba, Meguro-ku, Tokyo 153-8902, Japan
}%


\begin{abstract}
Cells generally convert external nutrient resources to support metabolism
and growth. Understanding the thermodynamic efficiency of this conversion is
essential to determine the general characteristics of cellular growth.
Using a simple protocell model with catalytic reaction dynamics to
synthesize the necessary enzyme and membrane components from nutrients, the entropy
production per unit cell-volume growth is calculated analytically and
numerically
based on the rate equation for chemical kinetics and
linear non-equilibrium thermodynamics.
The minimal entropy production per unit cell growth is found to be achieved at a
non-zero nutrient uptake rate, rather than at a quasi-static limit as in the
standard Carnot engine.
This difference appears because the equilibration mediated by the enzyme
exists
only within  cells that grow through enzyme and membrane synthesis.
Optimal nutrient uptake is also confirmed by
protocell models with many chemical components synthesized through a
catalytic reaction network. The possible relevance of the identified optimal
uptake to optimal yield for cellular growth is also discussed.
\end{abstract}

\maketitle


\section{INTRODUCTION}

A cell is a system that transforms nutrients into substrates for growth and division. By assuming that the nutrient flow from the outside of a cell is an energy and material source, the cell can be regarded as a system to transform energy and matter into cellular reproduction. It is important to thermodynamically study the efficiency of this transformation\cite{schrodinger,katzir,blackbox,westerhoff1982thermodynamics,skulachev1992laws}.

Regarding material transformation, the yield is defined as the molar concentration of nutrients (carbon sources) needed to synthesize a molar unit of biomass (cell content) and has been measured in several microbes \cite{pirt1965maintenance,Beeftink1990203,hao2010microbiological,russell1995energetics,russell1979}. As the conversion of nutrients to cell content is not perfect and material loss to the outside of a cell occurs as waste, the yield is generally lower than unity. The yield also changes with nutrient conditions, and measurements in several microbes show that the yield is maximized at a certain finite nutrient flow rate. The basic logic underlying the optimization of yield at a finite nutrient flow rate rather than at a quasi-static limit is not fully understood.

A cell can also be regarded as a type of thermodynamic engine to transform nutrient energy into cell contents. In this case, it is necessary to study the thermodynamic efficiency or entropy production during the process of cell reproduction.
The thermodynamic efficiency of metabolism has been measured in several microbes under several nutrient conditions \cite{westerhoff1983thermodynamic,westerhoff1983thermodynamic,rutgers1989thermodynamic,russell1995energetics,rutgers1989thermodynamic,von2006thermodynamics,thauer1977energy,teh2010thermodynamic}, and Westerhoff and others computed it by applying the phenomenological flow-force relationship of the linear thermodynamics to catabolism and anabolism \cite{westerhoff1982thermodynamics,westerhoff1982thermodynamics,rutgers1991control} to show that the efficiency is optimal at a finite nutrient flow. Although such a phenomenological approach is important for technological application, a physiochemical approach is also necessary to highlight difference between cellular machinery and the Carnot engine by characterizing the basic thermodynamic properties in a simple protocell model. Indeed, when viewed as a thermodynamic engine, a cell has remarkable differences from the standard Carnot-cycle engine.
The cell sits in a single reservoir, without a need to switch contacts between different baths. The cell grows autonomously to reproduce. To consider the nature of such a system, it is necessary to establish the following three points distinguishing the cell from the standard Carnot engine\cite{smith2008thermodynamics}.

First, cells contain catalysts (enzymes). The enzyme exists only within a compartmentalized cell encapsulated by a membrane and thus enables reactions to convert resources to intracellular components to occur within a reasonable time scale within a cell but not outside the cell. Without the catalyst, extensive time is required for the reaction. Thus, the reaction is regarded to occur only in the presence of the catalyst. This leads to an intriguing non-equilibrium situation: Let us consider the reaction $R+C \leftrightarrow P +C$ with $R$ as the resource, $P$ as the product, and $C$ as the catalyst. Then, under the existence of $C$, the system approaches an equilibrium concentration ratio with $[R]/[P] =\exp(-\beta(\mu_R-\mu_P))$ and $\mu_R$ and $\mu_P$ as the standard chemical potential of the resource and product, respectively, and with $\beta$ as the inverse temperature. In contrast, outside the cell, $R$ and $P$ are disconnected by reactions within the normal time scale; therefore, their concentration ratio can take on any value. In this sense, the external environment is non-equilibrium in nature, in contrast to the intracellular environment. This leads to a remarkable difference from the standard Carnot engine.

Second, while considering the dynamical process, it is important to note that the catalysts are synthesized within the cell as a result of catalytic reactions. The time scale to approach equilibrium can depend on the abundance of the catalyst, which depends on the reaction dynamics themselves. Based on the first and second points mentioned above, the approach to equilibrium in the intracellular environment depends on catalyst abundance, which also depends on the flow rate of nutrients from outside the cell. Hence, the thermodynamic efficiency could show non-trivial dependence upon the nutrient flow.

Third, cell volume growth results from membrane synthesis from nutrient components, facilitated by the catalyst, whereas the concentrations of catalyst and nutrient are diluted by cell growth, which results in a non-standard factor for thermodynamic characteristics.

These three issues, which are fundamental to cell reproduction, are mutually connected and thus inherent to a self-reproducing, or autopoietic, system. In contrast to dynamical systems studies for self-reproduction in catalytic reaction networks \cite{lancet,kaneko,kauffman,jain}, however, the thermodynamic characteristics for such systems have not been fully explored.\\
In the present study, we determine these characteristics using simple reaction dynamics consisting of the nutrient, catalyst, and membrane. In Sec. \rmtwo, we consider a simple protocell model consisting of a membrane precursor and catalyst under a given nutrient flow. The entropy production by chemical production per unit cell volume growth is shown to be minimized at a certain finite nutrient flow. The mechanism underlying this optimization is discussed in relation to the abovementioned three characteristics of a cell. The entropy production by material flow is discussed in Sec. \rmtwo.A and basically does not change the conclusion described above. A protocell model consisting of a variety of catalysts that form a network, together with nutrients and membrane precursors, has been investigated to confirm that the conclusion described above is not altered. The biological relevance of our results is discussed in Sec. \rmthr.

\section{ENTROPY PRODUCTION OF AN AUTOPOIETIC CELL}
\subsection{Two-component model}
\begin{figure}[htbp]
\begin{center}
  \includegraphics[width = 70 mm,bb = 0 0 592 412]{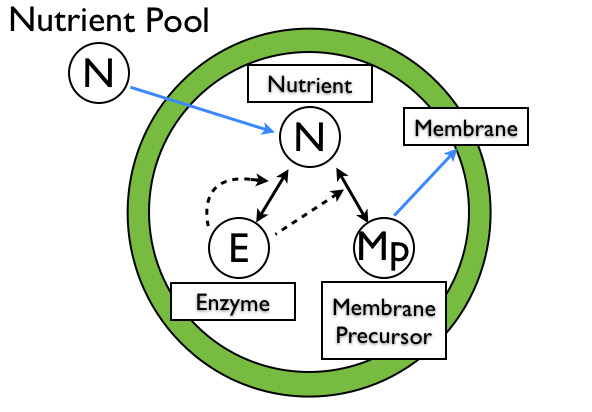}
\caption{ Schematic representation of our three-component protocell model. N, MP, and E denote nutrient, membrane precursor, and enzyme, respectively. The nutrient is taken up from the extracellular nutrient pool by diffusion, indicated by a blue arrow. All chemical reactions, indicated by black solid arrows, are reversible and catalyzed by the enzyme, as indicated by dashed arrows. Membrane precursors are transformed to the membrane as indicated by the green ring with some leaks. The membrane growth results in an increase in cell volume.
}
\label{fig:scheme}
\end{center}
\end{figure}

First, we study the entropy production $\sigma$ resulting from the intracellular reaction for the minimal protocell model consisting only of the synthesis of the enzyme and membrane precursor from the nutrient, which then leads to cellular growth \cite{klumpp2009growth,scott2010interdependence,scott2011bacterial,hao2010microbiological}(see FIG.\ref{fig:scheme} for schematic representation).
  The model consists of nutrient, membrane precursor, and enzyme, where the enzyme and membrane precursor are synthesized from the nutrient under catalysis by the enzyme. Moreover, by assuming that the diffusion constant of the nutrient is sufficiently large, the internal nutrient concentration is regarded to be equal to the external nutrient concentration. Based on the rate equation for chemical kinetics, our model is given by the following two-component ordinary differential equation
\begin{eqnarray}
  \frac{dx}{dt}&=&\kappa_x x( kX -x)-x\lambda, \nonumber \\
  \frac{dy}{dt}&=&\kappa_y x (lX-y)-\phi y-y\lambda. \label{eq:2}
\end{eqnarray}

where the variables $x$ and $y$ denote the concentrations of the enzyme and membrane precursor, respectively, whereas $\lambda \equiv \frac{1}{V}\frac{dV}{dt}$ denotes the cell volume growth rate to be determined.\\
Here, the notation of parameters is as follows:
\begin{itemize}
\item $X$\ :\ nutrient concentration.
\item $k={e}^{-\beta(\mu_x-\mu_{\mathrm{nut}})}$,\ $l=e^{-\beta(\mu_y-\mu_{\mathrm{nut}})}$, with $\mu_{\mathrm{nut}},\mu_x$ and $\mu_y$ as the standard chemical potential of nutrient, $x$, and $y$, respectively.
\item $\kappa_i$\ :\ catalytic capacity of the enzyme for $i$ component ($i=x,y$).
\item $\phi$\ :\ consumption rate of the membrane precursor to produce the membrane, such that the volume growth rate $\lambda$ is given by $\lambda = \gamma \phi y$, where $\gamma$ is the conversion rate from membrane molecules to cell volume.
\end{itemize}

In the stationary state, $\lambda$ takes a positive constant value of $y > 0$ for $X>0$ \footnote{In our model, another stationary solution $(x,y)=(0,0)$ exists. However this solution is an unstable fixed point of the differential equation for $X > 0$.}. Thus, the protocell volume increases exponentially in time. Here, we define the entropy production per unit volume at this steady growth state as $\sigma$. In computing $\sigma$, spatial inhomogeneity is not considered through the assumption of local homogeneous equilibrium. Thus, the entropy produced during the doubling in the protocell volume is given by
\begin{eqnarray*}
  S=\sigma \int_{0}^{T} V_0 e ^{\lambda t}dt
  =\frac{\sigma}{\lambda} V_0,
\end{eqnarray*}

where $V_0$ is the initial cell volume and $T$ is doubling time of the protocell volume.\\
\ \
We denote $\eta \equiv \sigma/\lambda$ as the entropy production per unit cell-volume growth. Generally, if $\eta$ is smaller, the thermodynamic efficiency for a cell growth is higher. For larger $\eta$, more energetic loss occurs in the reaction process. Hereafter, we study the dependence of $\eta$ on the nutrient condition and the growth rate $\lambda$.\\
In this subsection, we consider only the entropy production by the chemical reaction; the entropy production by the flow of chemicals from the outside of the cell will be considered in the next section.  The calculation of entropy production among different components is performed by virtually introducing chemical baths for different components that are mutually in disequilibrium and then applying linear non-equilibrium thermodynamics for calculation. This may result in stringent requisites; however, this step is adopted to address the thermodynamic efficiency of a cell with growth, as general steady-state thermodynamics are not established currently.
Then, the entropy production by the reactions is given by $\sigma = \sum_iJ_{i}\frac{A_i}{T}$, where $J_i$ is the chemical flow and $A_i$ is the affinity for each reaction. Here we set $T=1$ without losing generality.\\

\begin{figure}[!h]
  \begin{center}
    \includegraphics[width = 70 mm,bb=0 0 1440 1008,angle=0]{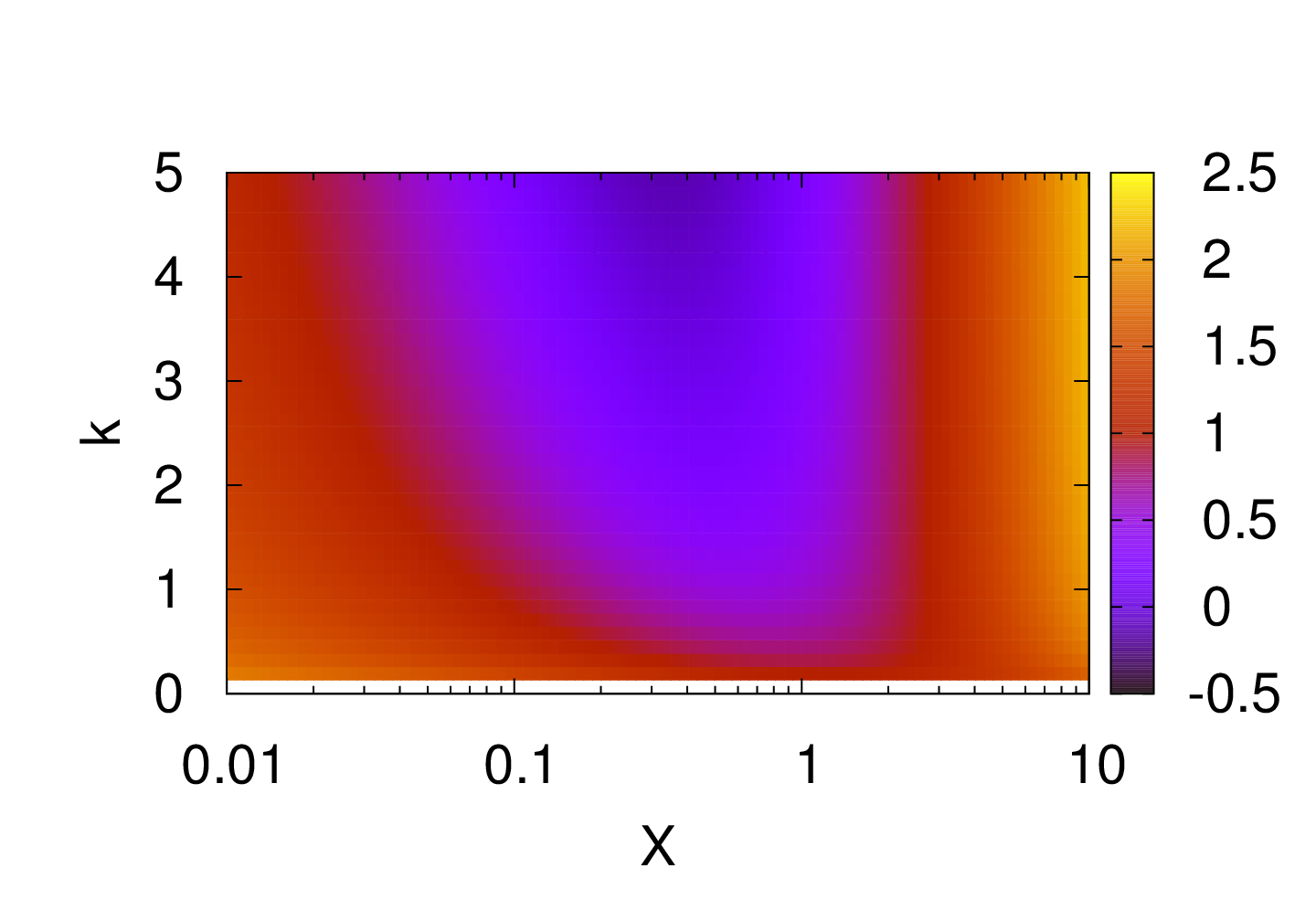}
    \caption{The logarithm of $\ti{\eta}$ plotted as a function of nutrient concentration and $\ti{k}$, with the color code given in the side bar. It is calculated from the solutions of Eq.(\ref{eq:tilde}). The parameter $\ti{\kappa}$ is chosen to be $1.0$. For given $\ti{k}$, there is an optimal nutrient concentration that gives the minimum $\eta$. (Tilde is omitted in the figure.)}
    \label{fig:eff_en_graph}
  \end{center}
\end{figure}

For calculation, we assume that $\kappa_x$ and $\kappa_y$ are identical for simplicity, denoted as $\kappa$. Then, by rescaling the variables as
\begin{eqnarray}
\ti{x}&=&x\gamma,\ \ \ti{y}=y\gamma,\ \ \nonumber \\ \ti{X}&=&lX\gamma,\ \ \tau = t\phi. \label{eq:scale}
\end{eqnarray}
Eq.(\ref{eq:2}) is written as
\begin{eqnarray}
  \frac{d\ti{x}}{d\tau}&=&\ti{\kappa}\ti{x}(\ti{k}\ti{X} -\ti{x})-\ti{x}\ti{y}, \nonumber \\
  \frac{d\ti{y}}{d\tau}&=&\ti{\kappa}\ti{x} (\ti{X}-\ti{y})-\ti{y}-\ti{y}^2 \label{eq:tilde},
\end{eqnarray}
where $\ti{\kappa} = \frac{\kappa}{\phi \gamma}$ and $\ti{k}=k/l$. The stationary solution of the equation for $\ti{\kappa}=1$ is given by

$$\ti{x}=\frac{\ti{k}\ti{X}(1+\ti{k}\ti{X})}{1+\ti{X}+\ti{k}\ti{X}},\ \ \ \ \ti{y}=\frac{\ti{k}\ti{X}^2}{1+\ti{X}+\ti{k}\ti{X}}.$$
Following this assumption, the entropy production by chemical reaction $\sigma$ at the stationary state is calculated as $\sigma=\sigma_x+\sigma_y$ with $\sigma_i=J_i\frac{A_i}{T}$ for the enzymatic reaction $i=x$ and for the membrane reaction $i=y$. Here, the flows are given by $\ti{J_x}=\ti{\kappa }\ti{x} (\ti{k} \ti{X}-\ti{x})$ and $\ti{J_y}=\ti{\kappa }\ti{x} (\ti{X}-\ti{y})$, whereas the affinities are given by $A_x=T\ln(\ti{k} \ti{X}/\ti{x})$ and $A_y=T\ln(\ti{X}/\ti{y})$. We omit the tilde for affinities because the affinities are not affected by scale transformation. Therefore, we obtain
\begin{eqnarray*}
\ti{\sigma} &=& \ti{\kappa }\ti{x} (\ti{k} \ti{X}-\ti{x})\ln(\ti{k} \ti{X}/\ti{x})+\ti{\kappa} \ti{x} (\ti{X}-\ti{y})\ln(\ti{X}/\ti{y}).
\end{eqnarray*}
The dependence of $\ti{\eta}\equiv \ti{\sigma}/\ti{y}=\gamma \eta$ upon $\ti{k}$ and $\ti{X}$, thus obtained, is plotted in FIG.\ref{fig:eff_en_graph} for $\ti{\kappa}=1$. As shown, the entropy production rate per unit growth shows a non-monotonic dependence on the nutrient concentration and is minimized at a non-zero nutrient concentration. Because nutrient uptake rate is a monotonic function of nutrient concentration, this result means that the entropy production rate per unit growth $\eta$ is minimal at a finite nutrient uptake rate. This result is in strong contrast with the thermal engine, where the entropy production is minimal at a quasi-static limit.

\begin{figure}[htbp]
  \subfigure[]{\includegraphics[width = 70 mm,bb = 0 0  1440 1008]{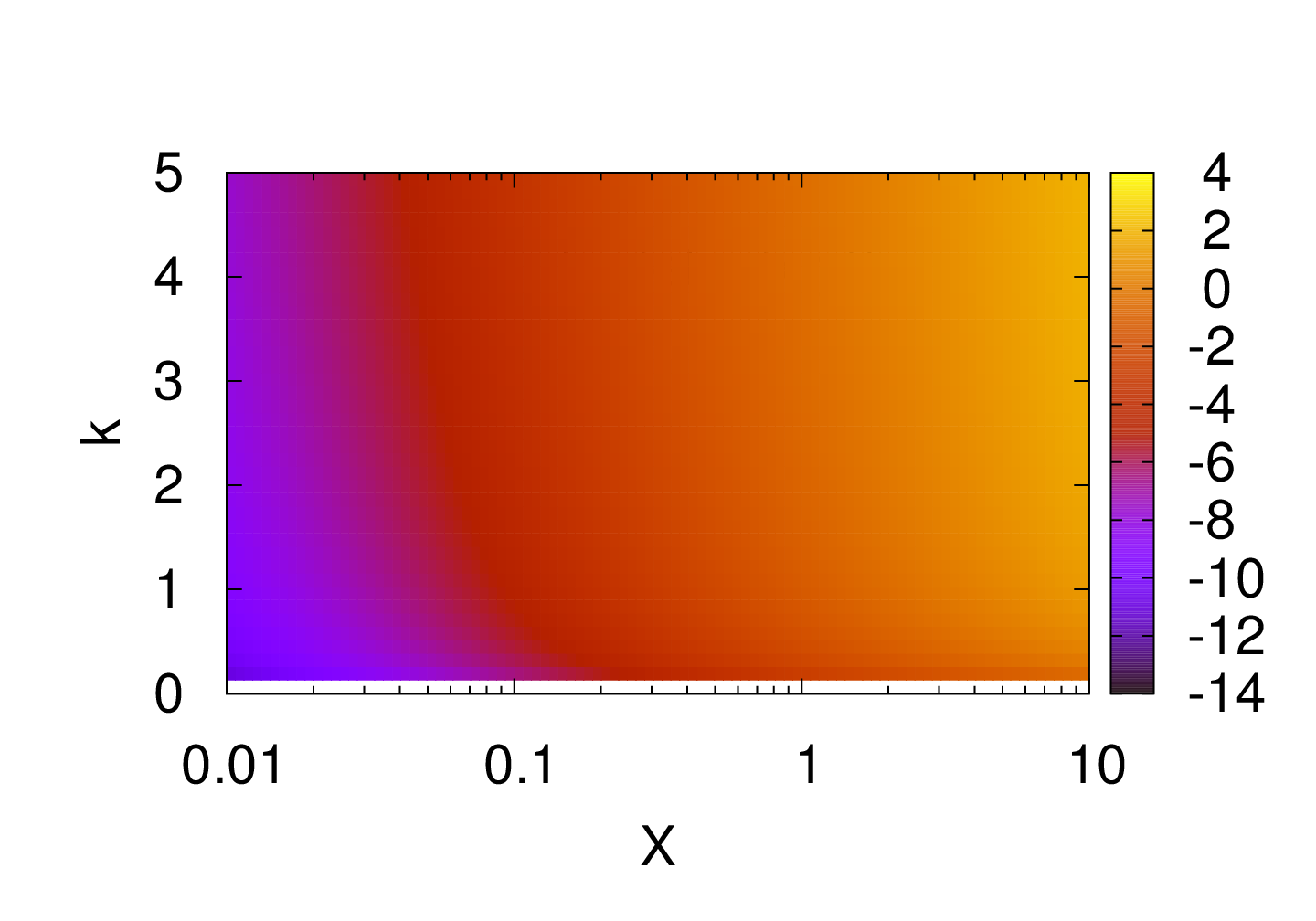}}
    \subfigure[]{\includegraphics[width = 70 mm,bb = 0 0 1440 1008]{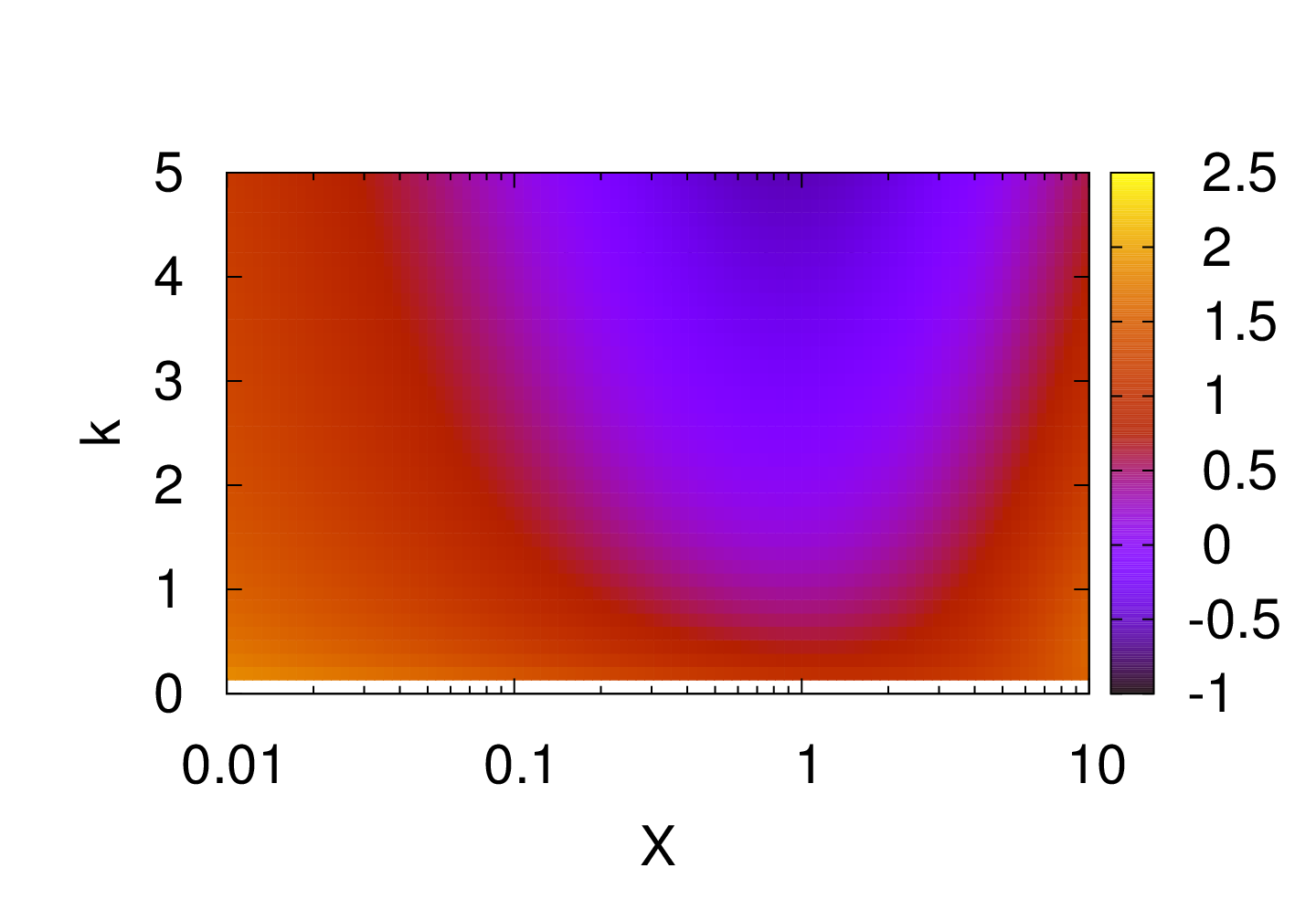}}
  \caption{The logarithm of the entropy production per unit growth rate $\sigma_x/\lambda$ and $\sigma_y/\lambda$ for the enzyme and membrane precursor synthesis reactions, respectively, plotted as a function of the nutrient concentration $\ti{X}$ and the rate constant $\ti{k}$, computed by Eq.(\ref{eq:tilde}). (a). $\sigma_x/\lambda$ for the enzyme producing reaction and (b) $\sigma_y/\lambda$ for the membrane precursor producing reaction.
}
  \label{fig:eff_en}
\end{figure}

FIG.\ref{fig:eff_en}(a),(b) shows the entropy production rate per unit growth $\sigma_x/\lambda,\sigma_y/\lambda$ for each reaction which produces component $x$ and $y$, respectively. This shows that the non-monotonic dependence on the nutrient in FIG.\ref{fig:eff_en_graph} is attributable to $\sigma_y/\lambda$.

As mentioned above, an important characteristic of cells is that intracellular reactions are facilitated by enzymes that are autonomously synthesized. Thus, the equilibrium distribution of chemicals in the presence of enzymes is different from the external chemical distribution. The decrease in $\eta$ under low nutrient concentrations is explained accordingly: The extracellular concentrations of the nutrient and of the membrane precursor are far from equilibrium in the presence of catalysts. Therefore, their intracellular concentrations under conditions of low nutrient uptake remain far from equilibrium and still similar to the external concentrations because of insufficiency of the enzyme. However, when the amount of nutrient uptake increases, the amount of enzyme increases and the system approaches intracellular equilibrium; therefore, the entropy production rate per unit growth decreases. \\
In contrast, with further increases in nutrient uptake, the entropy production rate increases as a result of the increase in cellular growth; entropy production $\sigma = \sum_iJ_i\frac{A_i}{T}$ by the reaction increases linearly with the reaction speed $J_i$. In the steady state, the reaction speed $J_i$ is roughly estimated by $\lambda x$, with $x$ as the concentration of the product of the reaction. For example, the dynamics of the enzyme concentration are given by $\frac{dx}{dt}=x( kX -x)-\lambda x$. At steady state, the enzyme production rate $x(kX -x)$ is balanced with $\lambda x$ according to Eq.(\ref{eq:2}). Thus, $\sigma_x$ increases with $\lambda x$. In summary, for a cell with a high growth rate, increased enzyme abundance is needed, which, however, leads to higher entropy production \footnote{For membrane production in Eq.(\ref{eq:2}), $(\phi+\lambda), y$ balances with the synthesis of the membrane precursor, but the tendency does not change.}\footnote{
Optimality at a finite flow rate was also discussed at the so-called weak-coupling regime in the linear non-equilibrium thermodynamics\cite{KedemCaplan}, where a linear relationship between the fluxes and forces,
$J_i=\Sigma_{j}L_{ij}X_j$ (with $1\leq i,j\leq2$) is adopted, in which $J_i$ is thermodynamic flow, and $X_i$ the conjugate thermodynamic force, and $L_{ij}$'s are the transport coefficients. Here, the degree of coupling is defined by $q=\frac{L_{12}}{\sqrt{L_{11}L_{22}}}=\frac{L_{21}}{\sqrt{L_{11}L_{22}}}$, while the thermodynamic efficiency is defined by the entropy production by the first process divided by the second process, given by $\eta'=-\frac{J_1X_1}{J_2X_2}.$  This efficiency  is known to take a local maximum at a finite flow, if the coupling is weak, i.e., $|q|<1$.  This optimality, however, is not related with the optimality discussed here. As discussed, ours is due to the equilibration under the existence of catalysts, whose synthesis speed is essential to give the optimality, in addition to the dilution by the cell-volume increase. Second, through straightforward calculation of thermodynamic variables for each reaction, it is shown that the optimality in our case is achieved under the tight-coupling regime i.e., $|q|=1$, where the optimality at a finite flow is not possible in the standard linear thermodynamics.
}.

In contrast, if the enzyme concentration is fixed externally, the entropy production rate per unit growth $\eta$ is minimized at the zero limit of nutrient concentration. In this case, the reaction dynamics Eq.(\ref{eq:2}) are reduced to
\begin{equation}
  \frac{dy}{dt}=c(lX-y)-\phi y-\phi y^2.
\end{equation}
where $c$ is a constant representing the concentration of the enzyme. In this case,
the stationary solution is given by $y=\frac{1}{2}[\ -(1+c/\phi)+\sqrt{(1+(c\phi)^2)+4clX/\phi}\ ]$, and accordingly ${\eta}^{-1}=(1+y)\ln(lX/y)$. There is no optimal nutrient concentration in this expression because $\frac{\partial {\eta}^{-1}}{\partial X}$ is always positive for any $X,l > 0$. This is consistent with the explanation mentioned above for Eq.(\ref{eq:tilde}). If the enzyme abundance is fixed to be independent of the nutrient uptake, the speed of approaching equilibrium is not altered by the nutrient condition; therefore, the entropy production just increases monotonically because of the cell volume growth.

\subsection{Additional entropy production by material flow}

Thus far, we considered only entropy production by chemical reactions. In addition, the material flow also contributes to entropy production, which is taken into account here.\\
To discuss the flow of nutrients, the dynamics of the nutrient concentration cannot be neglected. By including the temporal evolution of the nutrient concentration, the dynamics of the cellular state are given by
\begin{eqnarray}
\frac{ds}{dt}&=&-\kappa_x x( ks -x)-\kappa_y x(ls-y)\nonumber\\&-&s\lambda+D(s_{\mathrm{ext}}-s), \nonumber \\
  \frac{dx}{dt}&=&\kappa_x x (ks-x)-x\lambda \label{eq:3},\\
  \frac{dy}{dt}&=&\kappa_y x (ls-y)-\phi y -y \lambda. \nonumber
\end{eqnarray}
where $x,y$ and $s$ are the enzyme, membrane precursor, and nutrient concentration, respectively, and $\lambda=\frac{1}{V}\frac{dV}{dt}=\gamma \phi y$. The rate constants $k$ and $l$ are determined by the standard chemical potential of each chemical. Additionally, the nutrient is taken up with rate $D$ from the extracellular environment with a concentration $s_{\mathrm{ext}}$. \\
Entropy production by chemical flow is derived from nutrient uptake and membrane consumption, which (again by assuming linear nonequilibrium thermodynamics) are given by $\vec{J_{\rm s}}\cdot \nabla(-\mu_{\rm s}/T)$ and $\vec{J_{\rm y}}\cdot \nabla(-\mu_{\rm y}/T)$, respectively, where $\vec{J_{\rm i}}$ is the material flow of component $i$ and $\mu$ is the chemical potential. Integration of the terms over a narrow layer having a spatial gradient results in $D(s_{\rm ext}-s)\frac{s_{\rm ext}-s}{s}/T$ and $\phi y/T$. We neglect the entropy production of the solvent with the assumption that intra- and extracellular solvent concentrations are identical\footnote{The extracellular membrane concentration is assumed to be zero in our model; Eq.(\ref{eq:2}), we adopted entropy production of membrane consumption as a diffusion process.}. The contribution of dilution of the nutrient resulting from cellular growth is approximated as $\sigma_s\approx s\lambda$ by using the formula of entropy change resulting from the isothermal expansion of an ideal solution\footnote{Entropy production during isothermal expansion of an ideal solution from the initial volume $V_i$ to a terminal volume $V_t$ is $\Delta S=\ln(V_t/V_i)$ per unit mole. Because $\lambda$ is the volume expansion rate in this context and $V_t=V_i+\lambda \Delta t$, the change in entropy density is written as $\Delta s_v=\ln(1+\lambda \Delta t)$ per unit mole. The approximated formula is obtained by expanding $\ln(1+\lambda \Delta t)$ into the Taylor series and taking the limit of $\Delta t$ to zero.}; for other species, we use the same formula.\\

\begin{figure}[!h]
  \begin{center}
    \includegraphics[width = 70 mm,bb=0 0 720 504,angle=0]{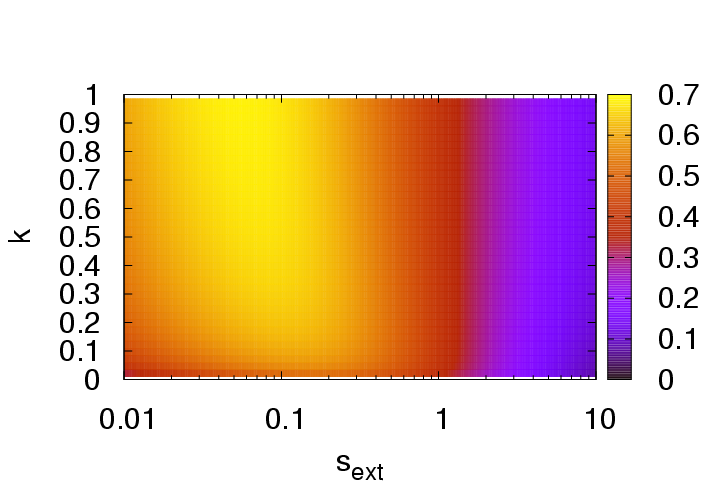}
    \caption{The thermodynamic efficiency for the model Eq.(\ref{eq:3}) plotted as a function of the external nutrient concentration $s_{\rm ext}$ and the rate constant $k$. The parameters were set as $\mu_s=0.0$, $D=1.0$, $\phi=1.0$, $\gamma=1.0$ and $\kappa_x=\kappa_y=1.0$. The standard concentrations were chosen to be $10^{-8}$.}
    \label{fig:thermodynamic_efficiency}
  \end{center}
\end{figure}

\begin{figure}[htbp]
  \subfigure[]{\includegraphics[width = 70 mm,bb = 0 0 720 504,angle=0]{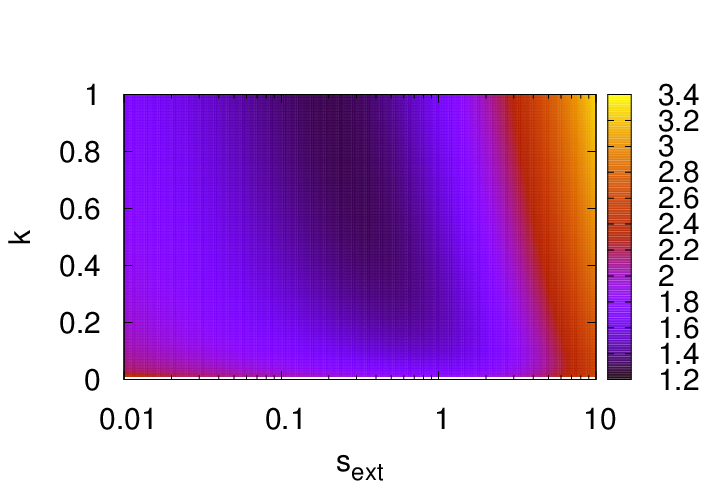}}
  \subfigure[]{\includegraphics[width=70 mm,bb = 0 0 720 504,angle=0]{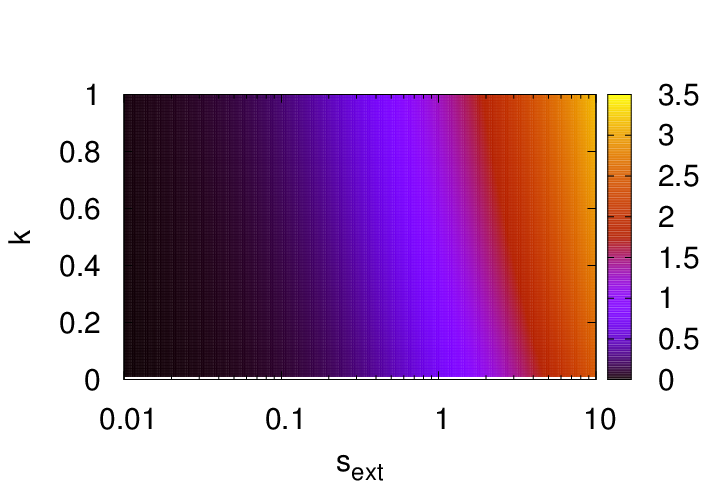}}
  \caption{The entropy production plotted as a function of the external nutrient concentration $s_{\rm ext}$ and the rate constant $k$, calculated from the fixed-point solution of Eq.(\ref{eq:3}); (a) the logarithm of total entropy production per unit cell growth, $\eta$; and (b) the logarithm of the entropy production per unit growth by material flow and dilution only. The parameters are chosen to be $\kappa_x=1.0,\ \kappa_y=1.0,\ D=1.0,\ \phi=1.0,\ \gamma=1.0$, and $l=k$.}
  \label{fig:eff_en_3nodes_diff}
\end{figure}

We choose that $\kappa_x,\kappa_y,D,\gamma$ and $\phi$ are equal to unity and that $\l=k$, for the sake of simplicity. Indeed, the characteristic behavior of $\eta$ is independent of this choice. Then, the fixed-point solutions of Eq.(\ref{eq:3}) are obtained against two parameters $k$ and $s_{\mathrm{ext}}$. From the solution, the entropy production per unit growth is computed, as shown in FIG.\ref{fig:eff_en_3nodes_diff}(a). We note that here again the minimal $\eta$ is achieved for a finite nutrient uptake, i.e., under nonequilibrium chemical flow. In FIG.\ref{fig:eff_en_3nodes_diff}(b), we plotted $\eta_{\mathrm{flow}}$, the entropy production excluding that derived from the chemical reaction. It increases monotonically with the external nutrient concentration. Entropy production is primarily derived from chemical reactions; therefore, the conclusion of subsection A is unchanged.\\
  Note that the so-called thermodynamic efficiency is defined as $\eta_{\rm th}=-\frac{J_a\Delta G_a}{J_c\Delta G_c}$ where $J_c$ and $J_a$ are the rates of catabolism and anabolism, and $\Delta G_c$ and $\Delta G_a$ are the affinities of catabolism and anabolism \cite{westerhoff1982thermodynamics,westerhoff1983thermodynamic}. Here, the optimality with regard to entropy production $\eta$ also leads to the optimal thermodynamic efficiency, which, in the present case, is computed by
$
\eta_{\rm th}=J_{\rm y}\mu_{\rm y}/J_{\rm s}\mu_{\rm s} 
$

where $J_{\rm s}=D(s_{\rm ext}-s)$ and $J_{\rm y}=\phi y$ are the absolute values of the uptake (and consumption) flow of chemical species $s$ (and $y$), and $\mu_{\rm i}$ is the chemical potential of the $i$th chemical species.  It is computed by using the chemical potential of nutrient $\mu_{\rm s}={\mu^0}_{\rm s}+T\ln(s/s_0)$ with ${\mu^0}_{\rm s}$ as the standard chemical potential for the nutrient and $s_0$ as its standard concentration (The chemical potential for $x$ and $y$ are computed in the same way). This thermodynamic efficiency also takes a local maximum value at a non-zero nutrient uptake rate (see FIG.\ref{fig:thermodynamic_efficiency}).\\

\section{EXTENSION TO A MULTI-COMPONENT MODEL}
\begin{figure}[htbp]
\begin{center}
  \subfigure[]{\includegraphics[width = 70 mm,bb=0 0 720 504]{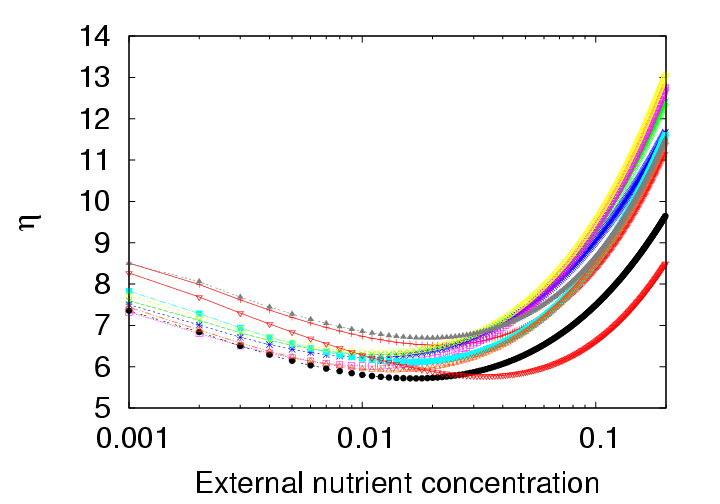}}
    \subfigure[]{\includegraphics[width = 70 mm,bb=0 0 720 504]{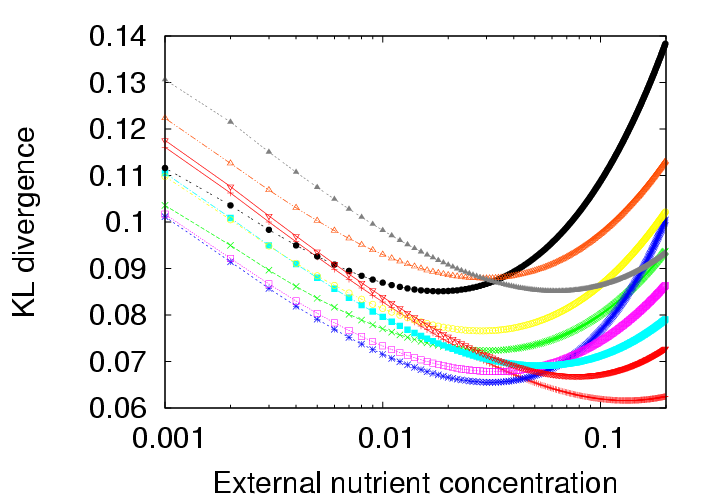}}
  \caption{The entropy production and deviation from equilibrium calculated from the steady-state solution of the multi-component model Eq.(\ref{eq:N_system}), plotted as a function of the external nutrient concentration. The results of 10 randomly generated networks are overlaid. (a).$\eta$; and (b). Kullback-Leibler divergence of the steady-state distribution from the Boltzmann distribution. The number of chemical species is set as $100$, whereas the parameter $\phi$ is chosen to be unity, and the ratio of the number of reactions to the number of chemical species is set to 3.}\label{fig:N}
  \end{center}
\end{figure}

It is worthwhile to check the generality of our result for a system with a large number of chemical species as in the present cell. For this purpose, we introduce a model given by
\begin{eqnarray}
 \frac{dx_1}{dt}&=&\sum_{\rm j=1}^{\rm N}\sum_{\rm k=2}^{\rm N-1}(C(1,j;k)k_{1j}x_j-C(j,1;k)k_{j1}x_1)x_k\nonumber \\&+&(X_1-x_1)-x_1\lambda, \nonumber \\
\frac{dx_i}{dt}&=&\sum_{\rm j=1}^{\rm N}\sum_{\rm k=2}^{\rm N-1}(C(i,j;k)k_{ij}x_j-C(j,i;k)k_{ji}x_i)x_k\nonumber \\&-&x_i\lambda ,\ \ \ (1<{\rm i}<{\rm N-1}), \label{eq:N_system}\\
\frac{dx_{N}}{dt}&=&\sum_{\rm j=1}^{\rm N}\sum_{\rm k=2}^{\rm N-1}(C(N,j;k)k_{Nj}x_j-C(j,N;k)k_{jN}x_{\rm N})x_k\nonumber \\&-&\phi x_{\rm N}-x_{\rm N}\lambda, \nonumber \\
\lambda&=& x_{N} \nonumber.
\end{eqnarray}
where the variables $x_1$, $x_{N}$, and $x_i\ (1<\rm{i}<\rm{N})$ denote the concentrations of the nutrient, membrane precursor, and enzymes, respectively, and $X_1$ is the external concentration of the nutrient. Each element of the reaction tensor $C(i,j;k)$ is unity if the reaction of $j$ to $i$ catalyzed by $k$ exists; otherwise, it is set to zero. Here, the nutrient and the membrane precursor cannot catalyze any reaction, whereas the other components $i=1,..N-1$ form a catalytic reaction network \cite{kaneko,kondo,awazu,furusawa}. All chemical reactions are reversible in our model; therefore $C(i,j;k)$ is equal to unity if and only if $C(j,i;k)$ equals unity. For the sake of simplicity, we assume that catalytic capacity, nutrient uptake rate, membrane precursor consumption rate, and the conversion rate from membrane molecule to cell volume are unity. The standard chemical potential $\mu_i$ for each chemical species is assigned by uniform random numbers within $[0,1]$, whereas $k_{ij}$ is given by ${\rm min}\{1,\exp(-\beta(\mu_i-\mu_j))\}$ accordingly \cite{awazu}.\\
Numerical simulations reveal that there again exists an optimal point of $\eta$ for each randomly generated reaction network of $N=100$. The dependence of $\eta$ on the nutrient concentration is plotted in FIG.\ref{fig:N}(a), overlaid for different networks. Although the nutrient concentration to give the optimal value is network-dependent, it always exists at a finite nutrient concentration; therefore, the entropy production is minimized at a non-zero nutrient concentration.

To determine a possible relationship with the optimality of $\eta$ and equilibrium in the presence of a catalyst
We also computed the Kullback-Leibler (KL) divergence of the steady state distribution from the equilibrium Boltzmann distribution as a function of the external nutrient concentration, expressed as
\begin{eqnarray*}
D_{KL}(p||q)=\sum_{i=1}^{N}p_i\ln\frac{p_i}{q_i},\ \ \\\Bigr( {\rm with}\ \ p_i=\frac{e^{-\mu_i}}{\sum_je^{-\mu_j}},\ \ q_i=\frac{{x^{st}}_i}{\sum_j{x^{st}}_j} \Bigl),
\end{eqnarray*}
where ${x^{st}}_i$ is the concentration of the $i$ th chemical species in the steady state. The KL divergence for each network shows non-monotonic behavior, as shown in FIG.\ref{fig:N}(b). Although the optimal nutrient concentration does not agree with the optimum for $\eta$, each KL divergence decreases in the region where $\eta$ is reduced. In this sense, it is suggested that the reduction of $\eta$ in our model Eq.(\ref{eq:N_system}) is related to the equilibration process of abundant enzymes synthesized as a result of a relatively high rate of nutrient uptake as discussed for Eq.(\ref{eq:2}) and Eq.(\ref{eq:3}).

\section{SUMMARY AND DISCUSSION}

To discuss the thermodynamic nature of a reproducing cell, we have studied simple protocell models in which nutrients are diffused from the extracellular environment and necessary enzymes for the intracellular reactions are synthesized to facilitate chemical reactions, including the synthesis of membrane components, which leads to the growth of cell volume. In the models, cell growth is achieved through nutrient consumption by the reactions described above. We computed $\eta$, which is the entropy production per unit cell volume growth and found that the value was minimized at a certain nutrient uptake rate. This optimization stems from the constraint that cells have to synthesize enzymes to facilitate chemical reactions, i.e., the autopoietic nature of cells. In general, the concentrations of nutrients and membrane components in extracellular environments are different from those in equilibrium achieved in the presence of enzymes, and the intracellular state moves towards equilibrium by synthesizing enzymes to increase the speed of chemical reactions. The equilibration reduces the entropy per unit chemical reaction. However, faster cell volume growth leads to a higher dilution of chemicals; therefore, faster chemical reactions are required to maintain the steady-state concentration of chemicals. Because entropy production by the reaction increases (roughly linearly) with the frequency of net chemical reactions, $\eta$ then increases for a higher growth range. Thus, the existence of an optimal nutrient content is explained by the requirement for reproduction mentioned in the introduction, i.e., equilibration of non-equilibrium environmental conditions facilitated by the enzyme, autocatalytic processes to synthesize the enzyme, and cell-volume increase resulting from membrane synthesis. \\
In the present model, all chemical components thus synthesized are not decomposed; they are only diluted. However, each component generally has a specific decomposition time or deactivation time as a catalyst. We can include these decomposition rates, which can also be regarded as diffusion to the extracellular environment with a null concentration. Then, the equilibration effect is clearer, although the results regarding optimal nutrient uptake are unchanged.\\
In the present study we focused on the entropy production that corresponds to dissipated energy per unit growth. In microbial biology, however, material loss is discussed as biological yield, as mentioned in the introduction, and it is thus reported that the optimal yield is achieved at a certain finite nutrient flow. Material loss is not directly included in the present model; therefore, we cannot discuss the yield derived directly from entropy production. However, it may be possible to assume that energy dissipation is correlated with material dissipation. \\
For example, the stoichiometry of metabolism is suggested to depend on dissipated energy \cite{insearch}.
Here, metabolism consists of two distinct parts: catabolism and anabolism. For catabolism, the energy is transported through energy currency molecules such as ATP, NADPH, and GTP, which are synthesized from the nutrient molecule. In this process, molecular decomposition also occurs, leading to the loss of nutrient molecules. In addition, the abundance of energy-currency molecules and the utilized energy are correlated. Hence, for both catabolism and anabolism, the energy dissipation and material loss are expected to be correlated. Indeed, a linear relationship between the yield and the inverse of thermodynamic loss (i.e., quantity similar to $1/\eta$ here) is suggested from microbial experiments \cite{insearch,roels1983energetics}.\\
Considering the correlation between energy and matter, the minimal entropy production at a finite nutrient flow that we have shown here may provide an explanation for the finding of optimal yield at a finite nutrient flow. Future studies should examine the relationship between minimal entropy production and optimal yield in the future by choosing an appropriate model that includes ATP synthesis and waste products in a cell. Currently, although our models are too simple to capture such complex biochemistry in a cell, they should initiate discussion regarding the thermodynamics of cellular growth.\\

\section*{ACKNOWLEDGMENTS}

The authors would like to thank A. Kamimura, N. Takeuchi, Y. Kondo,
T. Hatakeyama, A. Awazu, Y. Izumida, T. Sagawa, and T. Yomo for the useful discussion. The
present work
was partially supported by the
Platform for Dynamic Approaches to Living System from
the Ministry of Education, Culture, Sports, Science, and Technology of Japan and the Dynamical Micro-scale Reaction Environment Project of the Japan Science and Technology Agency.

\end{document}